\documentclass[reprint,prb,aps,superscriptaddress,amsmath]{revtex4-1}

\usepackage{graphicx}
\usepackage{amsmath}
\usepackage{txfonts}
\usepackage[usenames,dvipsnames,svgnames,table]{xcolor}

\begin{document}
\title{Nuclear energy gradients for internally contracted complete active space second-order perturbation theory: Multistate extensions}
\author{Bess Vlaisavljevich }
\author{Toru Shiozaki}
\email{shiozaki@northwestern.edu}
\affiliation{Department of Chemistry, Northwestern University, 2145 Sheridan Rd., Evanston, IL 60208}
\begin{abstract}
We report the development of the theory and computer program for analytical nuclear energy gradients
for (extended) multi-state complete active space perturbation theory (CASPT2) with full internal contraction.
The vertical shifts are also considered in this work.
This is an extension of the fully internally contracted CASPT2 nuclear gradient program, recently developed for a state-specific variant
by us [MacLeod and Shiozaki, J. Chem. Phys. {\bf 142}, 051103 (2015)];
in this extension, the so-called $\lambda$ equation is solved to account for the variation of the multi-state CASPT2 energies with respect to the
change in the amplitudes obtained in the preceding state-specific CASPT2 calculations, and the $Z$-vector equations are modified accordingly.
The program is parallelized using the MPI3 remote memory access protocol that allows us to perform efficient one-sided communication.
The optimized geometries of the ground and excited states of a copper corrole and benzophenone are presented as numerical examples.
The code is publicly available under the GNU General Public License.
\end{abstract}

\maketitle

\section{Introduction}
When the electronic structure is well described by a single Slater determinant, it can be modeled accurately by modern single-reference electron correlation methods.
For example, many-body perturbation and coupled-cluster theory have proven to be very effective, allowing researchers to predict energies, geometric parameters, and spectroscopic properties at times with accuracies on the order of wavenumbers.\cite{Bartlett2007RMP}
However, there are yet many chemical systems that exhibit an electronic structure requiring more than one electronic configuration to be taken into account.
This is particularly true for quasi-degenerate electronic states near avoided crossings or conical intersections in photochemical relaxation processes or in chemical reactions involving bond breaking.
These multi-configurational electronic structures are ubiquitous in inorganic chemistry, for instance,
in most transition metal compounds where states arising from varying the occupations of the {\it{d}}-orbitals are close in energy.

Multiconfigurational self-consistent field methods designed for such systems, for instance the complete active space self-consistent field (CASSCF) method,
are generalizations of the single-determinantal Hartree--Fock method.\cite{Helgakerbook}
The dynamical correlation models that use CASSCF wave functions as a reference have been developed over the past few decades;
the successful models include multi-reference second-order perturbation and configuration interaction methods.\cite{Werner1988JCP,Andersson1990JPC,Andersson1992JCP,Pulay2011IJQC,Szaley2012CR,Shiozaki2013MP} 
Among the most widely used forms of multi-reference second-order perturbation methods is the complete active space second order perturbation (CASPT2) approach
that employs the so-called internally contracted basis for efficiency.\cite{Andersson1990JPC,Andersson1992JCP,Pulay2011IJQC} 
The CASPT2 method has been applied to a variety of chemical problems including the spectra of organic compounds, transition metal complexes, and heavy elements.\cite{Roosbook,Gagliardi2007CSR} 

When multiple electronic states are nearly degenerate,
special care has to be taken, because errors in the state-specific approach may arise as the CASPT2 states are generally a linear combination of the CASSCF reference states
and may not be accurately approximated by a single CASSCF state.
The multi-state (MS) CASPT2 approach has been derived from quasi-degenerate perturbation theory based on the Bloch wave operator formalism,
which mixes different CASPT2 states by forming an effective Hamiltonian and diagonalizing it to obtain the mixing coefficients,\cite{Nakano1993JCP,Finley1998CPL}
Recently the ``extended" version of MS-CASPT2 (XMS-CASPT2) has been proposed\cite{Granovsky2011JCP,Shiozaki2011JCP3} 
to rectify its failure near conical intersections and avoided crossings where its potential energy surfaces are sometimes singular.
The XMS-CASPT2 method introduces a rotation of the reference CASSCF states to be used in the CASPT2 theory
such that the Fock operator is diagonal within the reference space and reduces to the standard MS-CASPT2 method when this rotation
is set to unit.

Implementation of analytical nuclear gradients for fully internally contracted CASPT2 has been an outstanding challenge for two decades.
The equations for the nuclear gradients of internally contracted multireference theories
are extremely complex, in part because the configuration interaction (CI) coefficients in the CASSCF reference wave functions are geometry dependent.
The nuclear gradient implementation was, therefore, limited to uncontracted\cite{Shepard1987IJQC,Shepard1992JCP,Nakano1998JCP,Dudley2003JCP,Theis2011JCP}
or partially contracted\cite{Celani2003JCP,Shiozaki2011JCP3,Gyorffy2013JCP} analogues until very recently,
when MacLeod and Shiozaki realized it for a fully internally contracted variant using an automatic code generation approach.\cite{MacLeod2015JCP} 
The use of full internal contraction in our program allows us to evaluate nuclear gradients far more efficiently than the existing implementations
based on uncontracted and partially contracted multireference perturbation theories.
For instance, the application of the analytical gradient program for a partially contracted variant by Celani and Werner\cite{Celani2003JCP} is limited to small molecules,
because it treats closed and active orbitals on an equal footing
[their implementation is technically limited to 32 occupied (i.e., closed and active) orbitals].
In the following, the efficiency of our code will be demonstrated by numerical examples that were intractable with previously reported methods.

Herein, we present the implementation of MS-CASPT2 and XMS-CASPT2 nuclear gradients using fully internally contracted basis functions.
Two full internal contraction schemes for (X)MS-CASPT2 wave functions, called the ``MS-MR" and ``SS-SR" schemes (see below), are considered.
Vertical shifts, which are often necessary in excited-state calculations, are included in the gradient code.\cite{Roos1995CPL}
The program is parallelized and ready for applications.
The ground- and excited-state geometries of a Cu corrole and benzophenone are presented to demonstrate the efficiency and versatility.

\section{Theory and Implementation}
In this section, we elaborate on the nuclear gradient theory for the fully internally contracted XMS-CASPT2 method. 
The working equations are based on the nuclear-gradient formulas for
the partially internally contracted XMS-CASPT2 reported by Shiozaki {\it et al.},\cite{Shiozaki2011JCP3}
which, in turn, was based on the CASPT2 Lagrangian proposed by Celani and Werner.\cite{Celani2003JCP}
Our implementation is interfaced to the {\sc bagel} program package\cite{bagel} as an extension to the state-specific CASPT2 nuclear gradient program recently reported by us.\cite{MacLeod2015JCP}
Hereafter, we use the following index notation: $i$ and $j$ label closed orbitals, $r$, $s$, and $t$ label active orbitals,
$a$ and $b$ label virtual orbitals, and $x$, $y$, $z$, and $w$ label general orbitals.

\subsection{Reference states}
The CASSCF reference states, labeled by $L$, $M$, and $N$, are a linear combination of Slater determinants $I$,
\begin{align}
|M\rangle = \sum_I c_{I,M} |I\rangle,
\end{align}
which are the eigen functions of the Hamiltonian within the active space, i.e.,
\begin{align}
\langle M| \hat{H} |N\rangle = \delta_{MN}E^\mathrm{ref}_M.
\end{align}
The orbitals are determined such that the state-averaged (SA) CASSCF energy, $E^\mathrm{ref,sa} = \sum_M W_M E^\mathrm{ref}_M$, is stationary
with respect to the orbital rotation parameters (i.e., the orbital-rotation gradients of $E^\mathrm{ref,sa}$, often labeled as $\mathbf{A}$, are zero).
The weight $W_N$ in the SA-CASSCF is set to be $1/N_\mathrm{st}$ in our program with $N_\mathrm{st}$ being the number of states included in the calculation.
It is also assumed in this work that the numbers of states included in the CASSCF and CASPT2 calculations are identical,
though this restriction can be easily lifted.

The XMS procedure\cite{Granovsky2011JCP,Shiozaki2011JCP3} has recently been proposed to make the MS-CASPT2 theory invariant with respect to rotations among reference states.
This invariance has been shown to be essential for accurate description of state crossings.\cite{Granovsky2011JCP,Shiozaki2013PCCP}
In XMS, the Fock operator is diagonalized within the reference configuration space,
\begin{align}
\sum_M \langle L | \hat{f} |M\rangle U_{MN} = U_{LN} \tilde{E}_N,
\end{align}
and, using this rotation matrix $U_{MN}$, we define a new set of reference configurations, 
\begin{align}
&|\tilde{M}\rangle = \sum_N |N\rangle U_{NM} = \sum_I \tilde{c}^M_I |I\rangle,\\
&\tilde{c}_{I,M} = \sum_N c_{I,N} U_{NM}.
\end{align}
Note that the Hamiltonian matrix within the reference space is no longer diagonal, 
\begin{align}
H^\mathrm{ref}_{MN} = \langle \tilde{M} | \hat{H} | \tilde{N}\rangle.
\end{align}
In the following, we present the equations for XMS-CASPT2 based on these rotated reference states.
The equations for MS-CASPT2 can be obtained by setting $U_{MN}$ to be a unit matrix.

\subsection{XMS-CASPT2 energy}
The first-order wave functions in the state-specific CASPT2 theory is parameterized using the internally contracted basis as
\begin{align} 
&|\Psi^{(1)}_L \rangle = \sum_N \hat{T}_{LN} |\tilde{N}\rangle = \sum_N \sum_\Omega \hat{E}_\Omega | \tilde{N}\rangle T_{\Omega,LN},\label{wfp}\\
&\hat{E}_\Omega = \left\{\hat{E}_{ai,bj},\,  \hat{E}_{ar,bi},\,  \hat{E}_{ar,bs},\,  \hat{E}_{ai,rj}, \right.\nonumber\\
&\quad \quad \left.\hat{E}_{ri,sj},\,  \hat{E}_{ar,st},\,  \hat{E}_{ri,st},\,  \hat{E}_{ai,rs},\, \hat{E}_{ar,si}\right\},
\label{class}
\end{align}
in which the internally contracted basis functions are generated from the union of the reference states.
This parametrization is often referred to as the ``MS-MR" contraction scheme.
Another popular parameterization used in the original MS-CASPT2 implementation by Finley {\it et al}.\cite{Finley1998CPL}
\begin{align}
&|\Psi^{(1)}_L \rangle = \hat{T}_{LL} |\tilde{L}\rangle = \sum_\Omega \hat{E}_\Omega | \tilde{L}\rangle T_{\Omega,LL},\label{wfp2}
\end{align}
is often referred to as the ``SS-SR" contraction scheme.
In what follows, we report the working equations based on the ``MS-MR" contraction primarily in terms of $\hat{T}_{LN}$ defined in Eq.~\eqref{wfp} to be consistent with the way our code is implemented.
All of the equations are implemented with the ``SS-SR" contraction as well; the working equation for the ``SS-SR" contraction is a subset of those with the ``MS-MR" contraction
and can be obtained by only retaining the diagonal terms.
Generally speaking, the ``MS-MR" contraction guarantees invariance of the theory with respect to any rotation among reference functions,\cite{Shiozaki2011JCP3}
whereas the ``SS-SR" contraction allows for the computation of MS-CASPT2 with computational costs that are linear in the number of states.\cite{Finley1998CPL}

The amplitudes $T_{\Omega,LN}$ for state $L$ are determined by the amplitude equation,
\begin{align}
\sum_{N}  \langle \tilde{M} |\hat{E}_\Omega^\dagger (\hat{f} - E_L^{(0)} + E_s)\hat{T}_{LN} | \tilde{N} \rangle
+ \langle \tilde{M} | \hat{E}_\Omega^\dagger \hat{H} | \tilde{L} \rangle = 0,\label{ampeq} 
\end{align}
which should hold for each $M$.
Here, the state-averaged Fock operator $\hat{f}$ is used, which is defined as 
\begin{align}
&\hat{f} = \sum_{xy} f_{xy} \{\hat{E}_{xy}\}_C = \sum_{xy} [\mathbf{h}+ \mathbf{g}(\mathbf{d}^\mathrm{(0)sa})]_{xy} \{\hat{E}_{xy}\}_C, \nonumber\\
&[\mathbf{g}(\mathbf{d})]_{xy} = \sum_{kl} \left[(xy|zw)d_{zw} - \frac{1}{4} (xw|zy)(d_{zw}+d_{wz})\right].
\end{align}
Normal ordering is used with respect to the closed-Fock vacuum and is denoted as $\{\cdots\}_C$:
\begin{align}
\{\hat{E}_{xy}\}_C = \left\{
\begin{array}{ll}
\displaystyle
2\delta_{xy} - \hat{E}_{xy}  & \quad x,y \in \{i\}  \\[4pt]
\hat{E}_{xy} & \quad \mathrm{otherwise}
\end{array}
\right.
\end{align}
The Hamiltonian $\hat{H}$ is also normal ordered with respect to the closed-Fock vacuum.
$\mathbf{d}^\mathrm{(0)sa}$ is the state-averaged zeroth-order density matrix, i.e.,
\begin{align}
d^\mathrm{(0)sa}_{xy} = \sum_N W_N \langle N| \hat{E}_{xy} |N\rangle = \sum_N W_N \langle \tilde{N}| \hat{E}_{xy} |\tilde{N}\rangle. 
\end{align}
$E_L^{(0)}$ in the amplitude equation is the zeroth-order energy of the $L$th state,
\begin{align}
E_L^{(0)} = \sum_{rs} \langle \tilde{L} | \hat{E}_{rs} | \tilde{L} \rangle f_{rs},
\end{align}
in which the summation runs only over active orbitals, owing to the normal ordering.
$E_s$ is a shift parameter.

The amplitude equation is solved using a subspace residual minimization algorithm. 
In general, the internally contracted basis is linearly dependent; therefore, we typically truncate the eigenvalues of the overlap matrix below 10$^{-9}$
when updating the amplitudes.
Once convergence is achieved, the effective Hamiltonian is constructed, 
\begin{align}
{H}^\mathrm{eff}_{LL'} &= H_{LL'}^{\mathrm{ref}}
+ \frac{1}{2}\sum_M \left(\langle \tilde{M} | \hat{T}_{LM}^\dagger \hat{H} | \tilde{L}' \rangle + \langle \tilde{L} | \hat{H} \hat{T}_{L'M} | \tilde{M}\rangle \right) \nonumber\\ 
&- \delta_{LL'}E_s \sum_{MN} \langle \tilde{M} | \hat{T}_{LM}^\dagger \hat{T}_{LN} | \tilde{N} \rangle,   
\end{align}
the diagonalization of which gives the XMS-CASPT2 energy (for $P$th state),
\begin{align}
\sum_M {H}^\mathrm{eff}_{LM} R_{MP} = R_{LP} E^\mathrm{MS}_{P}.
\end{align}
The final energy can be explicitly written as 
\begin{align}
E^\mathrm{MS}_{P} &= \sum_{MN} \langle \tilde{M} | \hat{H} | \tilde{N}\rangle R_{MP} R_{NP} 
+ \sum_{LMN} \langle \tilde{M} | \hat{T}_{LM}^\dagger \hat{H} | \tilde{N}\rangle R_{LP} R_{NP} \nonumber\\
&- E_s \sum_{LMN}\langle \tilde{M} | \hat{T}_{LM}^\dagger \hat{T}_{LN} | \tilde{N}\rangle R_{LP}^2.
\end{align}
In the subsequent subsections, we present an algorithm for computing the total derivatives of $E^\mathrm{MS}_{P}$ 
with respect to nuclear displacements.

\subsection{XMS-CASPT2 Lagrangian}
The XMS-CASPT2 energy $E^\mathrm{MS}_{P}$ is not stationary with respect to the
variation of the $T$ amplitudes. 
We therefore define the Lagrangian that is stationary with respect to the parameters in the CASPT2 part,
\begin{align}
\mathcal{L}_\mathrm{PT2} &= E^\mathrm{MS}_{P} + \sum_{LMN} \langle \tilde{M} | \hat{\lambda}_{LM}^\dagger (\hat{f} - E_L^{(0)} + E_s) \hat{T}_{LN} | \tilde{N} \rangle \nonumber\\
 &+  \sum_{LM} \langle \tilde{M} | \hat{\lambda}_{LM}^\dagger \hat{H} | \tilde{L}\rangle.
\label{lambdaeq}
\end{align}
in which the amplitude equations are included as a constraint. The $\lambda$ operator is defined similarly to the $T$ operator:
\begin{align}
\hat{\lambda}_{LN} |\tilde{N}\rangle = \sum_\Omega \hat{E}_\Omega | \tilde{N}\rangle \lambda_{\Omega,LN},
\end{align}
where $\lambda_{\Omega,LN}$ are the Lagrange multipliers for the corresponding amplitudes.
Note that taking the derivative of $\mathcal{L}_\mathrm{PT2}$ with respect to $\lambda_{\Omega,LM}$ yields the amplitude equation in Eq.~\eqref{ampeq}.
Taking the derivative of $\mathcal{L}_\mathrm{PT2}$ with respect to $T_{\Omega,LM}$, one obtains the so-called $\lambda$ equation:
\begin{align}
&R_{LP}\sum_N \langle \tilde{M} |\hat{E}_\Omega^\dagger  \hat{H} | \tilde{N}\rangle R_{NP}
- 2E_s R_{LP}^2 \sum_N \langle \tilde{M} | \hat{E}_\Omega^\dagger  \hat{T}_{LN} | \tilde{N} \rangle \nonumber\\
&+ \sum_{N}  \langle \tilde{M} | \hat{E}_\Omega^\dagger (\hat{f} - E_L^{(0)} + E_s)\hat{\lambda}_{LN} | \tilde{N} \rangle = 0.
\end{align}
With thus determined $\lambda$ amplitudes, $\mathcal{L}_\mathrm{PT2}$ is now stationary with respect to all of the $T$, $\lambda$, and $R$ parameters.

However, $\mathcal{L}_\mathrm{PT2}$ is not stationary with respect to orbital rotation or to the variation of the CI coefficients in the reference calculations.
By including the constraints associated with the CASSCF optimization,
we arrive at the total XMS-CASPT2 Lagrangian, which consists of $\mathcal{L}_\mathrm{PT2}$ and the constraints from the CASSCF procedure, i.e.,
\begin{align}
\mathcal{L} &= \mathcal{L}_\mathrm{PT2} + \frac{1}{2}\mathrm{Tr}[\mathbf{Z}(\mathbf{A} - \mathbf{A}^\dagger)]
-\frac{1}{2}\mathrm{Tr}[\mathbf{X}(\mathbf{C}^\dagger \mathbf{S}\mathbf{C}-1)] \nonumber\\
&+ \sum_N W_N \left[\sum_{I} z_{I,N} \langle I | \hat{H} - E^\mathrm{ref}_N | N\rangle  - \frac{1}{2}x_N (\langle N | N\rangle -1)  \right] \nonumber\\
&+ \sum_i^\mathrm{closed} \sum_j^\mathrm{frozen} z^c_{ij} f_{ij}
+ \sum_{MN} w_{MN} \langle \tilde{M}|\hat{f}|\tilde{N}\rangle.
\end{align}
Taking the derivatives of $\mathcal{L}$  with respect to $\mathbf{Z}$ and $z_{I,N}$, for instance, lead to
the orbital and CI conditions for the SA-CASSCF optimization.

\subsection{Input to the Z-vector equation}
Next we determine $\mathbf{Z}$, $z_{I,N}$ and $\mathbf{X}$ by solving the so-called $Z$-vector equation,\cite{Handy1984JCP,Celani2003JCP}
which is formally a set of linear equations corresponding to the stationary condition of the Lagrangian
with respect to orbital rotation parameters and variations of the CI coefficients. 
First, we calculate the derivatives of the CASPT2 Lagrangian $\mathcal{L}_\mathrm{PT2}$
with respect to the unitary generator for orbital rotations $\kappa_{xy}$ and the CI coefficients $c_{I,N}$, i.e.,
\begin{align}
&Y_{xy} = \frac{\partial \mathcal{L}_\mathrm{PT2}}{\partial \kappa_{xy}},\label{orbderiv}\\
&y_{I,N} = \frac{\partial \mathcal{L}_\mathrm{PT2}}{\partial c_{I,N}}.\label{cideriv}
\end{align}
The former is calculated from the correlated density matrices.\cite{Celani2003JCP}
The first-order contributions to the correlated density matrices are
\begin{align}
&d_{xy}^{(1)} = \sum_M \langle \tilde{M} | \hat{\mathcal{T}}_{PM}^\dagger \hat{E}_{xy}|P \rangle +  \sum_{MN} \langle \tilde{M} | \hat{\lambda}_{NM}^\dagger \hat{E}_{xy}|\tilde{N} \rangle, \\
&D_{xy,zw}^{(1)} = \sum_M \langle \tilde{M} | \hat{\mathcal{T}}_{PM}^\dagger \hat{E}_{xy,zw}|P \rangle +  \sum_{MN} \langle \tilde{M} | \hat{\lambda}_{NM}^\dagger \hat{E}_{xy,zw}|\tilde{N} \rangle, 
\end{align}
in which we have introduced the ``mixed" states,
\begin{align}
&|P\rangle = \sum_M |\tilde{M}\rangle R_{MP},\\
&\hat{\mathcal{T}}_{PN} = \sum_M \hat{T}_{MN} R_{MP}.
\end{align}
The mixed states have also been used in the previous implementation with partial internal contraction.\cite{Shiozaki2011JCP3} 
The second-order contributions are 
\begin{align}
&d_{xy}^{(2)} = \left\{ \begin{array}{ll}
\displaystyle \bar{d}_{xy}^{(2)} - \sum_L N_L \langle \tilde{L} | \hat{E}_{xy} | \tilde{L}\rangle & \quad x,y\in \{r\} \\[12pt]
\displaystyle \bar{d}_{xy}^{(2)} & \quad\mathrm{otherwise}
\end{array}
\right.
\end{align}
where $\bar{d}_{xy}^{(2)}$ and $N_L$ are respectively defined as
\begin{align}
&\bar{d}_{xy}^{(2)} = \sum_{LMN} \langle \tilde{M} | \hat{\lambda}_{LM}^\dagger \hat{E}_{xy}\hat{T}_{LN}|\tilde{N} \rangle,\\
& N_L = \sum_{MN}\langle \tilde{M} | \hat{\lambda}_{LM} \hat{T}_{LN}|\tilde{N}\rangle.
\end{align}
An efficient density-fitted algorithm for computing the orbital derivative $Y_{rs}$ using these correlated density matrices has been detailed in Ref.~\onlinecite{Gyorffy2013JCP}.

The derivatives of the CASPT2 Lagrangian $\mathcal{L}_\mathrm{PT2}$ with respect to $\tilde{c}_{I,N}$ are
\begin{align}
\tilde{y}_{I,M} &= 2 R_{MP} \langle I| \hat{H} | P\rangle + R_{MP} \sum_{N} \langle \tilde{N} | \hat{\mathcal{T}}^\dag_{PN} \hat{H} | I\rangle  \nonumber\\
& + \langle I | \hat{\mathcal{T}}^\dag_{PM} \hat{H} | P \rangle -2E_s \sum_{N}\langle \tilde{N} | \hat{\mathcal{T}}^\dag_{PN} \hat{\mathcal{T}}_{PM} | I \rangle  \nonumber\\
& +\sum_N \langle \tilde{N} | \hat{\lambda}^\dag_{MN} \hat{H} | I\rangle 
  +\sum_N \langle I | \hat{\lambda}^\dag_{NM} \hat{H} | \tilde{N}\rangle \nonumber\\ 
& + \sum_{LN} \langle \tilde{N} | \hat{\lambda}^\dag_{LN} (\hat{f}-E_L^{(0)}+E_s) \hat{T}_{LM} | I\rangle \nonumber\\
& + \sum_{LN} \langle I | \hat{\lambda}^\dag_{LM} (\hat{f}-E_L^{(0)}+E_s) \hat{T}_{LN} | \tilde{N}\rangle \nonumber\\
& + 2\sum_{rs} \langle I | \hat{E}_{rs} |\tilde{M} \rangle [W_M\mathbf{g}(\mathbf{d}^{(2)}) -  N_M \mathbf{f}]_{rs},
\end{align}
from which those with respect to $c_{I,N}$ [$y_{I,N}$ in Eq.~\eqref{cideriv}] can be easily calculated using the XMS rotation matrix $U_{rs}$ as
\begin{align}
y_{I,N} = \sum_M \tilde{y}_M U_{NM}.\label{cideriv2}
\end{align}
We have implemented a multi-passing algorithm for the computation of $\tilde{y}_{I,M}$, in which
the index $I$ is divided into several chunks such that the CI derivatives of the higher-order RDMs fit in distributed memory,
and the associated $\tilde{y}_{I,M}$ are computed separately.
This avoids the storage of the entire CI derivatives of the RDMs,
which previously limited the size of the active spaces.\cite{MacLeod2015JCP}

The Lagrange multiplier for the XMS constraint can be calculated using $\tilde{y}$ to be
\begin{align}
w_{MN} = -\frac{1}{2(E^{(0)}_M - E^{(0)}_N)} \sum_I(\tilde{c}_{I,M} \tilde{y}_{I,N} - \tilde{y}_{I,M} \tilde{c}_{I,N}),
\label{xmseq}
\end{align}
which follows from the chain rule.
The derivatives of the XMS term with respect to the orbital rotations and variation of CI coefficients are
\begin{align}
&Y'_{xy} = 2 \left[\mathbf{f}\check{\mathbf{d}} + \mathbf{g}(\check{\mathbf{d}}) \mathbf{d}^{(0)\mathrm{sa}}\right]_{xy},\\
&\tilde{y}'_{I,N} = 2\sum_M \langle I | \hat{f} | \tilde{M}\rangle w_{MN} + 2W_N \sum_{rs} \langle I | \hat{E}_{rs} |\tilde{N}\rangle [\mathbf{g}(\check{\mathbf{d}})]_{rs},
\end{align}
respectively, where $\mathbf{\check{d}}$ is an effective density matrix for the XMS term,
\begin{align}
{\check{d}}_{rs} = \sum_{MN} w_{MN}\langle \tilde{M} | \hat{E}_{rs} | \tilde{N}\rangle. 
\end{align} 
Note that $w_{MN}$ is set to zero when computing MS-CASPT2 nuclear gradients without the XMS procedure.

$Y_{xy}$ and $y_{I,N}$ are used to solve the $Z$ vector equation; the effective density matrices are then formed and contracted
to the gradient integrals. The details of the $Z$-CASSCF equations and the contraction to the gradient integrals using density fitting are found in Ref.~\onlinecite{Gyorffy2013JCP}
(see also Ref.~\onlinecite{Delcey2015JCP}). 

\subsection{Numerical stability analysis}
There are four steps in the XMS-CASPT2 energy and nuclear gradient evaluation, in which we solve a linear equation. Since 
solving linear equations could potentially introduce numerical instability, here we analyze the stability of these equations in detail.
These steps include the amplitude equation [Eq.~\eqref{ampeq}], the $\lambda$ equation [Eq.~\eqref{lambdaeq}], the XMS response term [Eq.~\eqref{xmseq}], and 
the $Z$-vector equation (Ref.~\onlinecite{Gyorffy2013JCP}).
The other steps are tensor multiplications, which are numerically stable.

First, the numerical instability in the amplitude equation (called the intruder-state problem) can be eliminated by the vertical shift $E_s$.\cite{Roos1995CPL}
This also means that the vertical shift can stabilize the $\lambda$ equation,
because the condition number of the $\lambda$ equation is identical to that of the amplitude equation
(the only difference between these two equations is in the source terms [see Eqs.~\eqref{ampeq} and \eqref{lambdaeq}]).

The response of the XMS rotation matrix [Eq.~\eqref{xmseq}] involves the division by the difference of the zeroth-order energies of each pair of reference states ($E^{(0)}_M-E^{(0)}_N$).
This apparent divergence reflects the fact that, when the zeroth-order energies are degenerate, the XMS procedure does not yield a unique set of reference states. 
However, when they are degenerate, it follows from quasi-degenerate perturbation theory\cite{Granovsky2011JCP}
that the final XMS-CASPT2 energies are invariant with respect to any rotations among reference states as well. Therefore, whenever $|E^{(0)}_M-E^{(0)}_N|$ is
very close to zero, one can simply set the corresponding element of the XMS response $w_{MN}$ to zero. 

Finally, we analyze the stability of the $Z$-vector equation. Let us use a short-hand notation $Ax = b$ for the $Z$-vector equation,
where $x$ is a column vector consisting of $Z_{xy}$ and $z_{I,N}$ and $b$ is that consisting of $Y_{xy}$ and $y_{I,N}$.\cite{Celani2003JCP,Gyorffy2013JCP}
The matrix $A$ on the left-hand side is the second derivatives of the SA-CASSCF energy with respect to the orbital rotations and variations of the CI coefficients. 
Since the SA-CASSCF solutions are minima in the parameter space, this matrix is guaranteed to be positive definite.
Whenever some of the eigenvalues of $A$ are close to zero, it means that the active space is ill-defined;
in other words, there are nearly empty or doubly-occupied orbitals in the active space,
making the state-averaged reference energy almost invariant with respect to these orbital rotations.
In practice, the $Z$-vector elements that correspond to these nearly redundant rotations can be set to zero.
It is important to note that the stability of the $Z$-vector equation is determined solely by the SA-CASSCF conditions (and not by the properties of electron correlation methods),
because all of the terms that are dependent on $\mathcal{L}_\mathrm{PT2}$ are contained in $b$ in the above abbreviated expression,
which is the source term of the linear equation. 

In summary, we have shown that, as long as the intruder-state problems in the amplitude equations are addressed by the vertical shift $E_s$,\cite{Roos1995CPL} there is no
additional numerical issue in the CASPT2 gradient evaluation.
Our experience with the code so far also supports this analysis.

\subsection{Code generator and parallelization}
All of the working equations are implemented using a code generator, named {\sc smith},
that we have recently reported.\cite{smith} Our code generator can handle spin-free equations for multireference electron correlation theories\cite{MacLeod2015JCP,smith}
and implements the following expressions:
\begin{subequations}
\begin{align}
&\langle N | \hat{E}^\dagger_\Omega \hat{G} \hat{A} | M\rangle \\ 
&\langle N | \hat{A'}^\dagger \hat{E}_\Lambda \hat{A} | M\rangle \\
&\langle I | \hat{A'}^\dagger \hat{G} \hat{A} | M\rangle
\end{align}
\end{subequations}
where $\hat{A}=\sum_\Omega A_\Omega \hat{E}_\Omega$ is an excitation operator, $\hat{G}=\sum_\Lambda G_\Lambda \hat{E}_\Lambda$ is a general operator (such as Hamiltonians),
$|N\rangle$ and $|M\rangle$ are arbitrary active-space configurations, and $I$ labels Slater determinants.
The generated code does not utilize symmetry at the moment.
It was previously used to implement analytical nuclear gradient programs for state-specific CASPT2.\cite{MacLeod2015JCP} 

In this work, we extensively make use of this machinery.
For instance, when evaluating Eq.~\eqref{ampeq}, the generated code is called within the three-fold nested loops over possible $L$, $M$, and $N$ states;
inside these loops, we set appropriate RDMs and energies to the tensors to be read by the generated code, 
\begin{align}
&\Gamma_{rs} = \langle \tilde{M}|\hat{E}_{rs}|\tilde{N}\rangle,\quad\Gamma_{rs,tu}=\langle \tilde{M}|\hat{E}_{rs,tu}|\tilde{N}\rangle,\quad\cdots,\nonumber\\
&T_\Omega = T_{\Omega, LN},\quad E^{(0)} = E^{(0)}_L. 
\end{align}
Using these quantities, the corresponding state-specific CASPT2 code is called to calculate the contribution to the residual vector.
The same approach is used in the $\lambda$ equation and the evaluation of the correlated density matrices and CI derivatives. 
All of the explicit formulas are complied in the supplementary information of Ref.~\onlinecite{MacLeod2015JCP}.

The programs generated by {\sc smith} are parallelized using the MPI remote memory access protocols (RMA).
The recently extended interfaces in the MPI3 standard
(e.g., {\tt MPI\_Rget}, {\tt MPI\_Rput}, and {\tt MPI\_Raccumulate})
allow us to check the completion of specific one-sided communications without global fencing.
At this moment, these one-sided communications are used as a blocking operation (i.e., we call these function and wait until their completion).
In tensor contractions, the data locality is used such that {\tt MPI\_Raccumulate} is intra-node while {\tt MPI\_Rget} can be inter-node; we made this decision because
{\tt MPI\_Raccumulate} is typically not implemented at the hardware level unlike {\tt MPI\_Rget} and {\tt MPI\_Rput}.
The index permutation of tensors is currently performed at the destination; further optimization using 
a scalable universal matrix multiplication algorithm (SUMMA)\cite{VanDeGeijin1997C,tiledarray}
to avoid the repeated permutation operations will be performed in the future.

\section{Numerical Examples} 
Optimizations of the ground and excited states of 
a copper corrole and benzophenone are presented.
Unless otherwise stated, all of the optimizations were performed using the {\sc bagel} package without imposing spatial symmetry.

\subsection{Optimization of S$_0$, S$_1$, and T$_0$ for a copper corrole}
\begin{figure}[tb]
\includegraphics[width=0.45\textwidth]{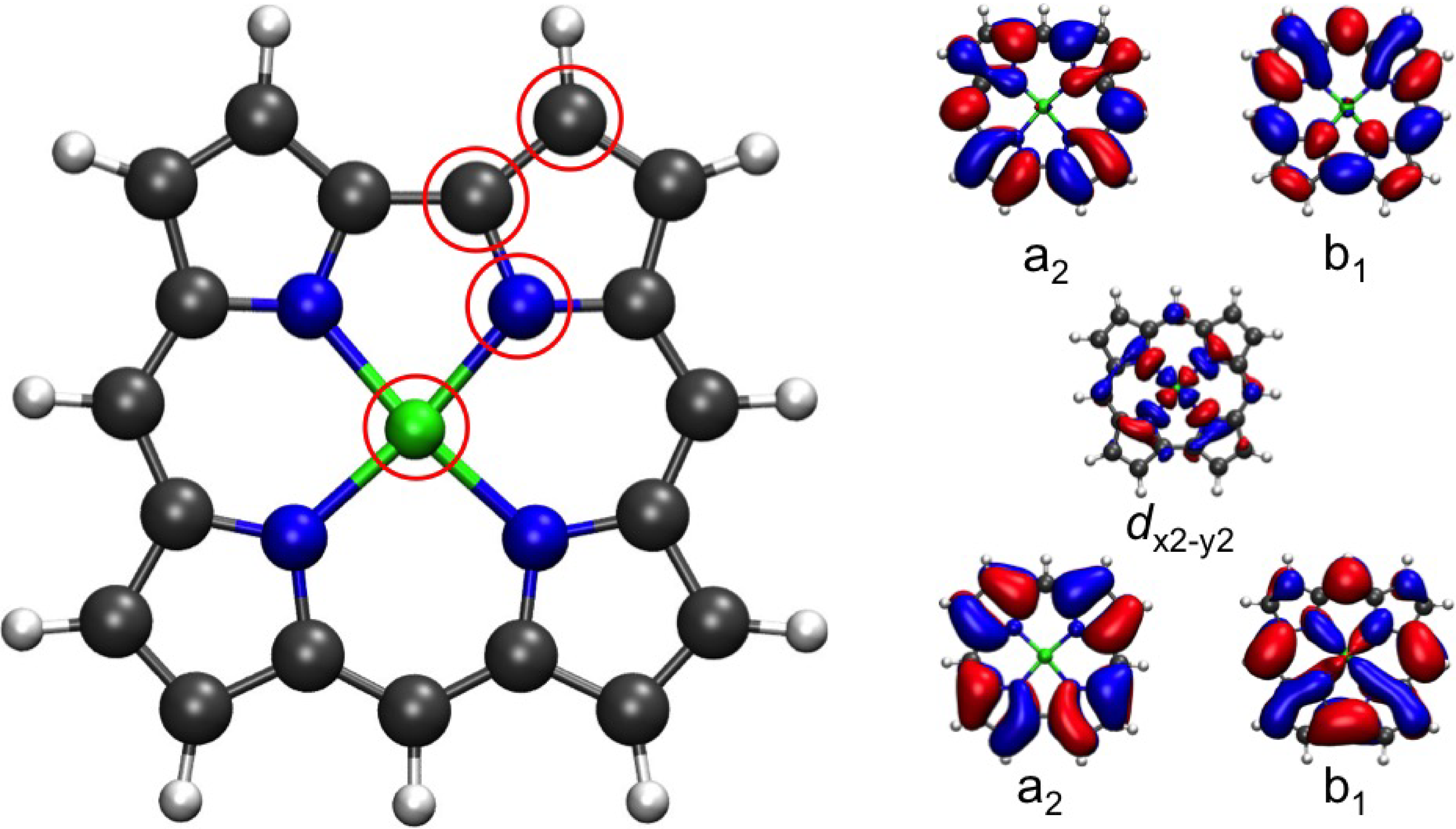}
\caption{\label{fig:corrole} The copper corrole complex. Active orbitals are listed on the right.
The resulting geometries range from planar (C$_{2v}$) to a saddled (C$_2$) structure.
}
\end{figure}

Corroles are non-innocent, porphyrin-like macrocycles that exhibit unique coordination chemistry and electronic structure.
While structurally similar to porphyrins, two of the pyrrole rings in the corrole are directly bound and the ligand is trianionic when fully deprotonated (see Fig.~\ref{fig:corrole}).
Corroles have an electronic structure that consists of the four so-called Gouterman orbitals,\cite{Gouterman1959JCP}
which include two a$_1$ and two b$_1$ orbitals where each has one occupied and one unoccupied orbital
(note that, although we label orbitals using the notation in the C$_{2v}$ symmetry, the actual symmetry of the molecule is C$_2$, not C$_{2v}$).
Metallocorroles have {\it{d}}-orbitals that are close in energy to the corrole $\pi$-system.
This, in combination with the non-innocent ligand, has been shown to yield a multiconfigurational ground state.
Furthermore, work by Pierloot {\it et al}. on the copper corrole suggested that as the geometry of the complex changes from planar to saddled,
the ordering of the two lowest singlet states inverts.\cite{Pierloot2010IC}
There has been much discussion in the literature regarding the formal oxidation states of the metal (and likewise the extent to which a radical is present on the ligand).\cite{Lemon2016ACIE}
The proposed electronic structures in the literature fall to two extremes:
(1) a d$^9$ Cu(II) center with a radical in the b$_1$ orbital of the corrole or (2) a closed shell Cu(III) center where no spin is present on the ligand.

In order to study how the geometry changes from the ground state singlet (S$_0$) to the first excited singlet (S$_1$), geometry optimizations were performed using XMS-CASPT2 
and the cc-pVDZ basis set along with the TZVPP JKFIT basis set (since there is no Cu basis set in cc-pVDZ JKFIT) for density fitting. A vertical shift of 0.2 was also employed.
The active orbitals include the four Gouterman orbitals\cite{Gouterman1959JCP} and the {\it{d}}$_{x2-y2}$ orbital resulting in a (4{\it{e}},5{\it{o}}) active space
(see Fig.~\ref{fig:corrole}).
The remaining four {\it{d}} orbitals are doubly occupied and sufficiently low in energy to remain in the inactive space.
The orbitals were state averaged over two states.

The largest contributions to the S$_0$ and S$_1$ wave functions are
(a$_2$)$^2$(b$_1$)$^1$(d$_{x2-y2}$)$^1$ and (a$_2$)$^1$(b$_1$)$^2$(d$_{x2-y2}$)$^1$, respectively.
The T$_0$ state was also optimized and yields a planer C$_{2v}$ structure.
Both of the two singlet states are formally Cu(II) and differ in energy by 8.7 kcal/mol at the C$_{2v}$ T$_0$ XMS-CASPT2 optimized geometry.
The geometry change between the minima of these two states is most notable in a saddling motion that can be quantified by averaging the C--C--N--Cu dihedral angle (denoted $\Omega$)
where the first carbon is a $\beta$-carbon on the pyrrole group (see red circles in Fig.~\ref{fig:corrole}).
When starting from a geometry with $\Omega$ = 2$^{\circ}$, the convergence to the minimum geometry of the S$_0$ state took 31 geometry optimization steps.
For this optimized structure, $\Omega$ is very distorted at 18.5$^{\circ}$. On the other hand the nearly planar $\Omega$ = 2$^{\circ}$ structure was a much better guess
for the S$_1$ state which converged in 15 geometry steps to a structure with $\Omega$ = 0.1$^{\circ}$.
When the ``SS-SR" scheme is employed, the value for $\Omega$ at the S$_0$ and S$_1$ minima changes only slightly yielding values of 0.5$^{\circ}$ and 18.3$^{\circ}$, respectively.
Each geometry optimization step took 41 minutes using the ``MS-MR" scheme and 20 minutes using the ``SS-SR" scheme using 128 CPU cores (Xeon E5-2650 2.0~GHz).
Optimized geometries and their corresponding total energies are available in the Supporting Information.

We note in passing that this system was selected partly because Pierloot {\it et al}.
reported that the MS-CASPT2 method overestimated the coupling between these states.\cite{Pierloot2010IC}
In our calculations, however, the coupling between CASPT2 states was found to be small.

\subsection{Optimization of the S$_0$ and S$_1$ states of benzophenone}

\begin{figure}[tb]
\centering
\includegraphics[width=0.45\textwidth]{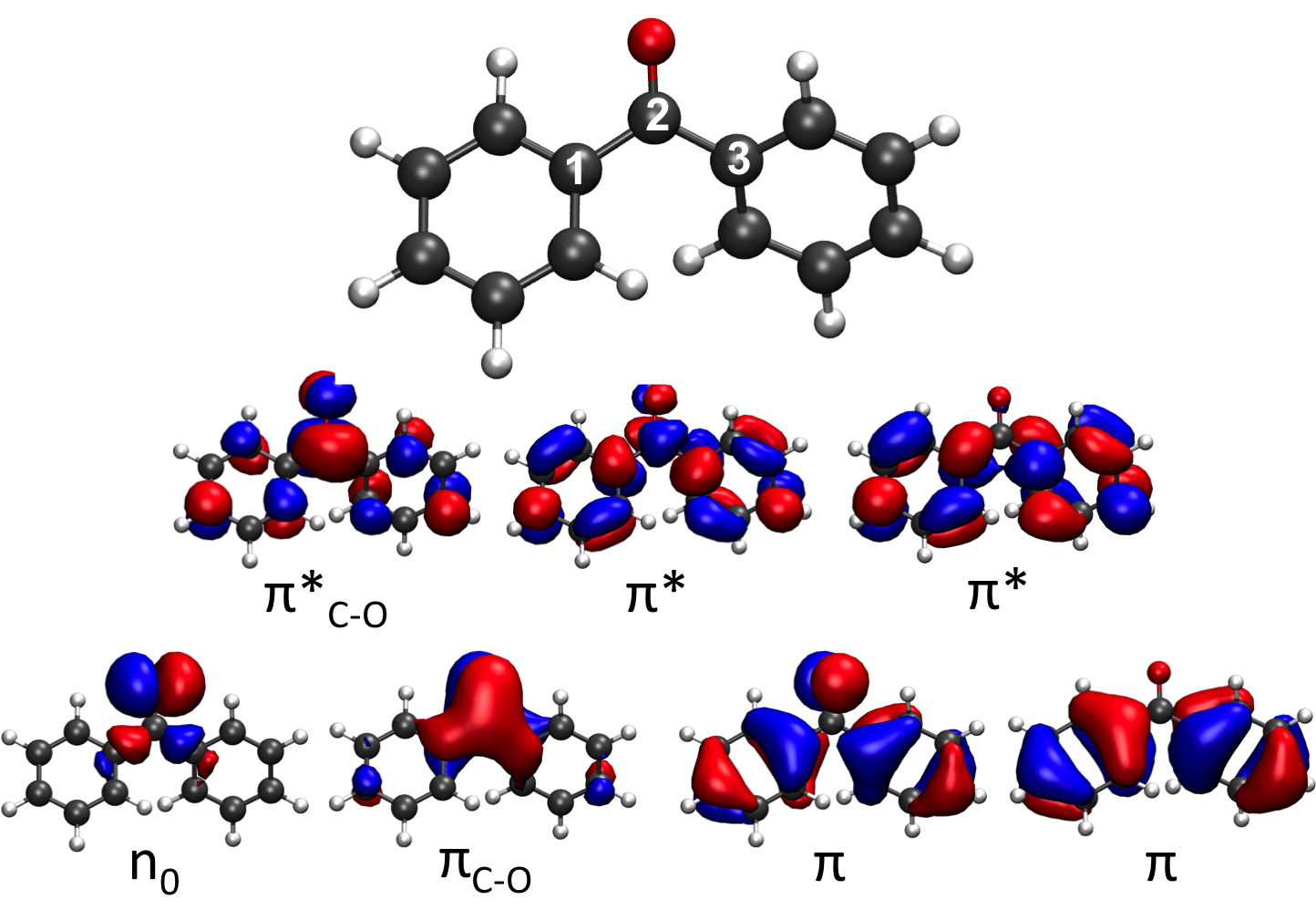}

\caption{\label{fig:benzophenone} The benzophenone molecule and the active orbitals included in the (8{\it{e}},7{\it{o}}) active space.}
\vspace{-0.2in}
\end{figure}

The benzophenone molecule is widely used as a triplet sensitizer because it rapidly undergoes intersystem crossing from the excited (S$_1$) state
to triplet states with a quantum yield close to one.
This transition has been observed in single crystals, in solution, and in isolated matrices.\cite{Katoh1997CPL,Aloise2008JPCA,Ohmori1988JPC}
The most pronounced change in geometry between the S$_0$ to S$_1$ states involves the elongation of the C-O bond and a change in the tilting angle between the two phenyl rings
(see Fig.~\ref{fig:benzophenone} and Table~\ref{table:benzophenone}).\cite{Hoffman1970JPC,Sergentu2014PCCP}
The initial S$_0$ to S$_1$ excitation and the nature of the subsequent transition into the T$_1$ state was recently studied in detail by Sergentu {\it{et al.}}\cite{Sergentu2014PCCP}
Here, we present the XMS-CASPT2 optimized geometry for benzophenone in the ground and first excited singlet states.
The first excitation in benzophenone includes a transition from the non-bonding oxygen lone pair orbital (n$_0$) to the $\pi$* orbitals in the C--O double bond.
To optimize the geometry of the $S_1$ state, we constructed an (8$e$,7$o$) active space consisting of n$_0$ and several $\pi$ orbitals.
The active orbitals are shown in Fig.~\ref{fig:benzophenone}. 

\begin{table}[tb]
\caption{\label{table:benzophenone} Select geometric parameters for the optimized S$_0$ and S$_1$ states. Ph--Ph refers to the angle between the two planes of the phenyl rings.
$\Delta E$ is the adiabatic excitation energy in eV.
}
\begin{ruledtabular}
\begin{tabular}{cccc}
& CASPT2 & CASSCF &  PBE0 \\
\hline
C2--O          &  1.23 / 1.35  &  1.21 / 1.37  &  1.21 / 1.30  \\
C1--C2        &  1.50 / 1.44  & 1.51 / 1.44  &   1.49 / 1.44 \\
C2--C3        &  1.50 / 1.45  & 1.53 / 1.44   &   1.49  / 1.44 \\
C1--C2--C3  &  119 / 129    &  120 / 131  &   120  / 127 \\
Ph--Ph        &  53 / 39        &  56 / 37 &   54 / 40 \\
$\Delta E$ & 3.36 & 3.45 & 3.66  
\end{tabular}
\end{ruledtabular}
\end{table}

The geometry was optimized using XMS-CASPT2 with the cc-pVDZ basis set and the corresponding JKFIT basis employing the ``SS-SR" scheme and a vertical shift of 0.2.
The first three singlet states were included.
For comparison, the S$_0$ and S$_1$ states were also optimized with SA-CASSCF (8{\it{e}},7{\it{o}}) and PBE0 using the {\sc molcas} and {\sc turbomole} packages,
respectively.\cite{molcasref,Furche2014WIREs}
In the SA-CASSCF calculation, the same basis set and state-averaging was used with Cholesky decomposition.
PBE0 calculations were performed using the def2-TZVP basis set.
Select geometric parameters from these optimizations are given in Table~\ref{table:benzophenone}, and the coordinates are given in the Supporting Information.

The ground state has a Ph--Ph angle of 53$^{\circ}$ and a C--O bond distance of 1.23 \AA.
In the S$_1$ state, the $\pi$* orbital on the C--O bond is occupied and as a result the C--O bond distance increases to 1.35 \AA.
For all three methods, the bond distances demonstrate the same trend; however, the C1--C2 and C2--C3 bond distances differ slightly more in the S$_0$ state
with CASSCF than with either CASPT2 or PBE0. Additionally, the difference between the S$_0$ and S$_1$ Ph--Ph angles is 14$^\circ$ when either CASPT2 or PBE0 are used;
however, the difference with CASSCF is larger by 5$^\circ$.
One geometry step took about 19 min using 64 CPU cores (Xeon E5-2650 2.0~GHz).
Efficient computation of analytical gradients presented here is a first step toward reliable simulations of photochemical dynamics,
for which standard DFT fails to describe state crossings between the ground and excited states.

\section{Conclusions}
We have developed an efficient program for calculating analytical nuclear energy gradients for MS- and XMS-CASPT2 with full internal contraction,
by extending the state-specific CASPT2 analytical gradient code previously reported by us.
We derived and implemented the so-called $\lambda$ and $Z$-vector equations based on the (X)MS-CASPT2 Lagrangian.
The program is parallelized using the MPI3 RMA instructions. 
We demonstrated the efficiency of the code by optimizing a copper corrole complex and benzophenone for their S$_0$ and S$_1$ states.  
The program is interfaced to the {\sc bagel} program package,\cite{bagel} which is publicly available under the GNU General Public License. 
The code generator {\sc smith3} is also open source.\cite{smith}
We will investigate in the near future the use of symmetry and scalable parallel algorithms
by extending the code generator and interfacing the code with parallel runtime libraries.

\begin{acknowledgments}
Jeff Hammond is thanked for helpful comments on the MPI3 RMA.
This work has been supported by the Department of Energy, Basic Energy Sciences (Grant No.~DE-FG02-13ER16398) and the Air Force Office of Scientific Research Young Investigator Program (Grant No.~FA9550-15-1-0031).
\end{acknowledgments}

\end{document}